\documentclass[reqno,12pt]{amsart}
\usepackage{amsmath}
\usepackage{amsxtra}
\usepackage{amscd}
\usepackage{amsthm}
\usepackage{amsfonts}
\usepackage{amssymb}
\usepackage{eucal}
\usepackage[matrix,arrow,curve]{xy}
\usepackage{cite}
\usepackage{mathtools}
\usepackage[hypertex]{hyperref}

\theoremstyle{plain}
\newtheorem{Thm}[subsection]{Theorem}
\newtheorem{Cor}[subsection]{Corollary}
\newtheorem{Lem}[subsection]{Lemma}
\newtheorem{Prop}[subsection]{Proposition}
\newtheorem{Conj}[subsection]{Conjecture}

\theoremstyle{definition}
\newtheorem{Def}[subsection]{Definition}

\theoremstyle{remark}

\newtheorem{Rem}[subsection]{Remark}

\newtheorem{conj}{Conjecture}

\numberwithin{equation}{section}
\renewcommand{\rm}{\normalshape}

\newcommand{\refl}[1]{Lemma ~\ref{L:#1}}

\newcommand{\bref}[1]{\textbf{\ref{#1}}}

\newenvironment{thm}[1]%
    { \begin{Thm} \label{T:#1}  \ifShowLabels \TeXref{T:#1} \fi }%
    { \end{Thm} }

\renewcommand{\th}[1]{\begin{thm}{#1} \sl }
\renewcommand{\eth}{\end{thm} }

\newenvironment{lemma}[1]%
    { \begin{Lem} \label{L:#1}  \ifShowLabels \TeXref{L:#1} \fi }%
    { \end{Lem} }
\newcommand{\lem}[1]{\begin{lemma}{#1} \sl}
\newcommand{\elem}{\end{lemma}}

\newenvironment{propos}[1]%
    { \begin{Prop} \label{P:#1}  \ifShowLabels \TeXref{P:#1} \fi }%
    { \end{Prop} }
\newcommand{\prop}[1]{\begin{propos}{#1}\sl }
\newcommand{\eprop}{\end{propos}}

\newenvironment{corol}[1]%
    { \begin{Cor} \label{C:#1}  \ifShowLabels \TeXref{C:#1} \fi }%
    { \end{Cor} }
\newcommand{\cor}[1]{\begin{corol}{#1} \sl }
\newcommand{\ecor}{\end{corol}}

\newenvironment{defeni}[1]%
    { \begin{Def} \label{D:#1}  \ifShowLabels \TeXref{D:#1} \fi }%
    { \end{Def} }
\newcommand{\defe}[1]{\begin{defeni}{#1} \sl }
\newcommand{\edefe}{\end{defeni}}

\newenvironment{remark}[1]%
    { \begin{Rem} \label{R:#1}  \ifShowLabels \TeXref{R:#1} \fi }%
    { \end{Rem} }
\newcommand{\rem}[1]{\begin{remark}{#1}}
\newcommand{\erem}{\end{remark}}

\newenvironment{conjec}[1]%
    { \begin{Conj} \label{Co:#1}  \ifShowLabels \TeXref{Co:#1} \fi }%
    { \end{Conj} }
\renewcommand{\conj}[1]{\begin{conjec}{#1} \sl }
\newcommand{\econj}{\end{conjec}}

\newcommand{\eq}[1]%
    { \ifShowLabels \TeXref{E:#1} \fi
       \begin{equation} \label{E:#1} }
\newcommand{\eeq}{ \end{equation} }

\newcommand{\prf}{ \begin{proof} }
\newcommand{\epr}{ \end{proof} }

\def\be{\begin{equation}}
\def\ee{\end{equation}}
\def\ba{\begin{array}}
\def\ea{\end{array}}

\newif\ifShowLabels
\ShowLabelstrue
\newdimen\theight
\def\TeXref#1{%
    \leavevmode\vadjust{\setbox0=\hbox{{\tt
        \quad\quad  {\small \rm #1}}}%
    \theight=\ht0
    \advance\theight by \lineskip
    \kern -\theight \vbox to
    \theight{\rightline{\rlap{\box0}}%
    \vss}%
    }}%
\ShowLabelsfalse

\numberwithin{equation}{section} \makeatletter
\@addtoreset{equation}{section}

\makeatletter
\g@addto@macro{\endabstract}{\@setabstract}
\newcommand{\authorfootnotes}{\renewcommand\thefootnote{\@fnsymbol\c@footnote}}%
\makeatother

\begin{document}

\begin{flushright}
FIAN-TD-2014-18 \\
\end{flushright}

\bigskip
\bigskip
\begin{center}
{ \large\textbf{ Correlation Functions in Unitary Minimal Liouville
\\ \vspace{3mm}Gravity  and Frobenius Manifolds}}
\vspace{10mm}
 \par
 \vspace{.8cm}

  \normalsize
  \authorfootnotes
V. Belavin\footnote{E-mail: belavin@lpi.ru}\textsuperscript{1,2}
\par
\bigskip

\begin{tabular}{ll}
$\qquad ^{1}$~\parbox[t]{0.9\textwidth}{\normalsize\raggedright
\small{I.~E. Tamm Department of Theoretical Physics, \,\,P.~N. Lebedev Physical \\Institute, Leninsky prospect 53, 119991 Moscow, Russia}}\\
$\qquad^{2}$~\parbox[t]{0.9\textwidth}{\normalsize\raggedright
\small{Department of Quantum Physics, Institute for Information Transmission Problems, Bolshoy Karetny per. 19, 127994 Moscow, Russia}}\\
\end{tabular}\\
\end{center}
\vspace{.8cm}
\begin{abstract}
We continue to study minimal Liouville gravity (MLG) using a dual approach
based on the idea that the MLG partition function is related to the tau
function of the $A_q$ integrable hierarchy via the resonance
transformations, which are in turn fixed by conformal selection rules. One
of the main problems in this approach is to choose the solution of the
Douglas string equation that is relevant for MLG. The appropriate solution
was recently found using connection with the Frobenius manifolds. We use
this solution to investigate three- and four-point correlators in the
unitary MLG models. We find an agreement with the results of the original
approach in the region of the parameters where both methods are applicable.
In addition, we find that only part of the selection rules can be satisfied
using the resonance transformations. The physical meaning  of the nonzero
correlators, which before coupling to Liouville gravity are forbidden by the
selection rules, and also the modification of the dual formulation that
takes this effect into account remains to be found.
\end{abstract}
 \bigskip

\section{Introduction}
\label{sec:Introduction}

The two-dimensional (2D) theory of Liouville gravity  arises in the context
of noncritical string theory \cite{Polyakov:1981rd}. This theory is called
minimal Liouville gravity (MLG) in the case where the target space in the
string sigma action or, equivalently, the matter sector in the worldsheet
theory  is represented by some minimal CFT model. Being a conformal theory,
MLG can be studied by the standard methods of 2D CFT \cite{Belavin:1984vu}
(see, e.g., \cite{ThreePoint,Yang-Lee,BelAlZam,MLGTorus}). 
There is another, {\it dual} approach based on the natural
geometric interpretation of MLG models as theories describing renormgroup
fixed points of  2D quantum systems on fluctuating surfaces.

Historically, the idea of fluctuating geometries first led to the development of the matrix-model
(MM) approach to 2D gravity \cite{Kazakov:1985ea,Kazakov:1986hu,
Kazakov:1989bc,Staudacher:1989fy,Brezin:1990rb,Douglas:1989ve, Gross:1989vs,
Douglas:1989dd}. The coincidence of the spectra of gravitational dimensions
in MLG and MM \cite{Knizhnik:1988ak} gave reason to believe that the two
approaches describe one theory. Unfortunately, the results of the MM and MLG
approaches do not coincide on the level of correlation functions
\cite{Moore:1991ir}; therefore, the MM approach is in fact concerned with
some other class of 2D gravity models.

It is nevertheless believed that MLG models can also be described by tools
similar to those in MM approach. In particular, the  formalism based on the
Douglas string equation \cite{Douglas:1989dd}, which arises in the MM
context, can be applied to MLG with some modifications. In this paper, we
call this approach to MLG the {\it dual} approach to distinguish it from
what is traditionally called the MM approach to 2D gravity.

It was first pointed out in \cite{Moore:1991ir} that a possible modification
is related to the ambiguity in adding contact term interactions when
defining MLG. Indeed, these contact terms are not controlled in the
corresponding CFT theory and should be fixed {\it by hand} in order to
define the integrated correlators. This ambiguity must be taken into account
to reconcile the results of the dual and initial continuous approaches on
the level of the correlation functions. Technically,  this effect leads to
the possible mixing of the Liouville coupling constants having compatible
gravitational dimensions. This is called resonance transformations. The
natural idea, proposed and further developed in \cite{Moore:1991ir},
\cite{Belavin:2008kv}, is that this freedom must be fixed by MLG selection
rules. On the sphere the selection rules contain the following requirements: vacuum
expectation values of physical operators are absent, the two-point
correlators are diagonal, and the conformal fusion rules are satiafied.
In \cite{Belavin:2008kv}, this idea was applied to  the series $M_{2,2p+1}$
of MLG models, and the explicit form of the resonance transformation in
terms of Legendre polynomials was found.

The new progress in developing the dual approach to MLG
\cite{Belavin:2013,VBelavin:2014fs,BB:2014} is due to the connection
between the Douglas string equation, ingtegrable Gelfand--Dikij
hierarchies, and so-called $A_q$ Frobenius manifolds. This connection was
used in \cite{VBelavin:2014fs} to analyze the unitary series $M_{q,q+1}$
of minimal models coupled to Liouville gravity. Based on properties of  $A_q$ Frobenius
manifolds, it was suggested that flat coordinates on the Frobenius manifold
is most appropriate for analyzing the correlation functions. In particular,
it was shown that the special solution of the Douglas  string equation that
is relevant for MLG has a simple form in flat coordinates. This idea was
verified on the level of one- and two-point correlation numbers on the
sphere,\footnote{See \cite{Spodyneiko} for some results of applying the dual
approach to MLG in the torus case.} and it was shown that only using this
solution allows satisfy the basic requirements of the method that the
so-called selection rules inherited by MLG from the conformal fusion rules
of the CFT model in its matter sector be satisfied.

This paper is organized as follows. In Section \bref{sec:Duglas}, we briefly
review the dual approach to MLG and describe its connection with $A_q$
Frobenious manifolds. Section \bref{sec:Reson} is devoted to analyzing the
resonance transformations. Sections \bref{sec:onepoint}, \bref{sec:twopoint},
\bref{sec:threepoint}, and \bref{sec:fourpoint} are respectively concerned
with computations the one-, two-, three-, and four-point correlation
functions. Section \bref{sec:Concl} contains concluding remarks. Some
computation details are presented in Appendices \bref{sec:StrConst},
\bref{sec:Jacobi}, and \bref{sec:App4point}.

\section{Dual approach to MLG and Frobenius manifold structure}
\label{sec:Duglas}

In this paper, we restrict our attention to the series of unitary models
$M_{q,q+1}$ coupled to Liouville gravity in the spherical topology. In this
case, the approach is formulated as follows. We introduce the so-called
action $S$, which depends on $q-1$ parameters $u^1,u^2,\dots,u^{q-1}$,
\be \label{action ambigious}
S[u^{\alpha}]=\underset{y=\infty}{\text{res}}
\bigg (Q^{\frac{2q+1}{q}}(y)+\!\!\!\!\sum_{1\leq n\leq m \leq q-1}\!\!\!\! t_{mn} (\mu,\{\lambda_{kl}\})\,\,Q^{\frac{(q+1) m-q n}{q}}(y) \bigg)\,,
\ee
where the polynomial
\be\label{Q}
Q(y)=y^q+u^{q-1} y^{q-2}+u^{q-2} y^{q-3}+...+u^{1}y^0\,.
\ee
The set of $\lambda_{kl}$ denotes the Liouville couplings, and the functions
$t_{mn} (\mu,\lambda)$ are defined by the resonance transformations
discussed in more detail below. The Duglas string equation \cite{Douglas:1989dd}
at genus zero \cite{Ginsparg:1990zc} has the form
\be \label{action principle}
\frac{\partial S}{\partial u^\alpha} = 0\,, \qquad \alpha=1,...,q-1\,.
\ee
The main claim of the approach is that among the solutions of this system,
there exists s special solution ${\bf u}_*=(u^1_*,u^2_*,\dots,u^{q-1}_*)$
that can be used to construct the generating function of the correlators in
MLG.

It was shown in \cite{Belavin:2013} that the parameters $u^\alpha$ can be
interpreted as the coordinates on the $(q-1)$-dimensional Frobenius manifold
such that the metric in these coordinates is given by
\be \label{metricu0}
\bigg(\frac{\partial}{\partial u^\alpha},\frac{\partial}{\partial u^\beta}\bigg) =-\underset{y=\infty}{\text{res}}
\frac{\frac{\partial Q(y)}{\partial u^\alpha}
\frac{\partial Q(y)}{\partial u^\beta}}{\frac{dQ}{dy}}\,.
\ee
To define the structure of the Frobenius manifold, we associated its points
$\bf u$ with the $(q-1)$-dimensional Frobenius algebras $A_q({\bf u})$
(depending on the parameters $u^\alpha$) such that \eqref{metricu0}
represents a pairing of the algebra elements with the invariance property
that $(ab,c)=(a,bc)$ for arbitrary elements $a,b,c$ of the algebra. We
recall that a finite-dimensional commutative and associative algebra with
unity is called a Frobenius algebra if such an additional invariant pairing
is defined for its elements. In the context of MLG, we deal with the $A_q$
Frobenius algebra that is the algebra of polynomials modulo the ideal
generated by $Q'(y)$. Using definition \eqref{metricu0}, all  necessary
properties of the Frobenius manifold (such as flatness of the metric,
existence of the Frobenius potential, etc.)~can be checked in the
{\it initial} coordinates  $u^\alpha$ \cite{Dubrovin:1992dz}. This
interpretation of the parameters  $u^\alpha$ turns out to be very efficient.
It was found in \cite{VBelavin:2014fs} that the problem of choosing the
relevant solution can be solved by changing from the initial $\bf u$ to the
flat coordinates ${\bf v}({\bf u})$. More precisely, properties of the
Frobenius manifolds were used to show that the relevant solution of the
string equation ${\bf v}_*$ in the flat coordinates becomes
${\bf v}_*^{(0)}=(v_{*1},0,0,\dots)$ in the limit where the Liouville
coupling constants are equal to zero. This solution is unique, i.e., only
this solution gives the generating function (defined below) for which the
necessary selection rules  are satisfied on the level of one- and two-point
correlators: it gives zero vacuum expectation values of the physical fields
(except unity), and the two-point correlators are diagonal.

\subsection{Flat coordinates}
It follows from the properties of the Frobenius manifold that there exists a
one-parametric deformation of the flat connection defined as follows
\cite{Dubrovin:1992dz}. The deformed coordinates\footnote{The deformed
coordinates are flat coordinates with respect to the deformed connection
defined below.} are given by $\theta_\alpha(z)=\sum_{k=0}^{\infty} z^k \theta_{\alpha,k}$
such that $\theta_\alpha(0)=v_\alpha$. The transformation from the initial
coordinates to the flat coordinates is defined by
\begin{equation}
\theta_{\alpha,k}=-c_{\alpha,k} \underset{y=\infty}{\text{res}} Q^{k+\frac{\alpha}{q}}(y)\,,
\label{theta}
\end{equation}
where
\begin{equation}
c_{\alpha,k}^{-1}=\bigg(\frac{\alpha}{q}\bigg)_{k+1}
\end{equation}
and $(a)_n=\Gamma(a+n)/\Gamma(a)$ is the Pochhammer symbol. The flatness
condition is equivalent to the recurrence relation
\begin{equation}
\frac{\partial^2 \theta_{\lambda}(z)}{\partial v^{\alpha}\partial v^{\beta}}=z C^{\gamma}_{\alpha\beta}
\frac{\partial \theta_{\lambda}(z)}{\partial v^{\gamma}}\,.
\label{Recursion0}
\end{equation}
The corresponding deformation of the Levi-Civita connection is given by the
modification of the Christoffel symbols
\be
\Gamma_{\alpha\beta}^{\gamma}\rightarrow
\Gamma_{\alpha\beta}^{\gamma}+z C_{\alpha\beta}^{\gamma}\,.
\ee
The metric in the flat coordinates has the simple form
\begin{equation}
\eta_{\alpha\beta}=-q\underset{y=\infty}{\text{res}}\frac{\frac{\partial Q(y)}{\partial v^\alpha}
\frac{\partial Q(y)}{\partial v^\beta}}{Q'(y)}=
\delta_{\alpha+\beta,q}\,.
\end{equation}
In particular, $v^{\alpha}=v_{q-\alpha}$.

It is highly nontrivial that deformed coordinates \eqref{theta} can be
regarded as local densities of the commuting Hamiltonians of the integrable
Gelfand--Dikij hierarchies. An important consequence of this fact
\cite{Belavin:2013} is that the MLG generating function is just the
logarithm of the tau function with the simple representation
\begin{equation}
Z=\frac 12 \int_0^{\bf{v}_*} C_{\alpha}^{\beta\gamma} \frac{\partial S}{\partial v^{\beta}}
 \frac{\partial S}{\partial v^{\gamma}} d v^{\alpha}\,.
\label{Z}
\end{equation}
Here, the upper limit $\bf{v}_*$ is the appropriate solution of the Douglas
string equation
\be
\frac{\partial S(\bf{v}_*)}{\partial v^{\alpha}}=0\,, \qquad \alpha=1,\dots,q-1\,.
\label{streqflat}
\ee
The integral in \eqref{Z} is independent of the integration contour, which
means that the integrand is a closed one-form. The structure constants of
the Frobenius algebra in the flat coordinates $C_{\alpha}^{\beta\gamma}=
C_{\alpha,q-\beta,q-\gamma}$ here are
\begin{equation}
C_{\alpha\beta\gamma}=-q\underset{y=\infty}{\text{res}}\frac{\frac{\partial Q(y)}{\partial v^\alpha}
\frac{\partial Q(y)}{\partial v^\beta}\frac{\partial Q(y)}{\partial v^\gamma}}{Q'(y)}\,.
\end{equation}
In particular, because $C_{1\alpha\beta}=\eta_{\alpha\beta}$, we obtain
\begin{equation}
C_{1}^{\alpha\beta}=\delta_{\alpha+\beta,q} \qquad \text{and}
\qquad C_{\alpha}^{q-1,\beta}=\delta_{\alpha,\beta}\,.
\label{C_properties}
\end{equation}
It follows from the definition of the Frobenius manifold that there exists a
function $F({\bf v})$ such that
\begin{equation}\label{triple}
C_{\alpha\beta\gamma}({\bf v}) =\frac{\partial^3 F({\bf v})}{\partial v^{\alpha} \partial v^{\beta} \partial v^{\gamma}}\,, \quad \alpha, \, \beta, \, \gamma=1, \dots, n.
\end{equation}

Before discussing the correlation functions, the following remark is in
order. The structure constants in the flat coordinates are currently
unknown, but we find that with the  form of generating function \eqref{Z}
and the properties of the relevant solution ${\bf v}_*$ taken into account,
the general expression for the structure constant is not needed for
calculating the correlation function. Instead, we need the coefficients of
the expansion
\be
C_{\alpha\beta\gamma}(v_1,v_2,v_3,\dots)=C_{\alpha\beta\gamma}(v_1,0,0,\dots)+
\sum_{\rho=1}^{q-1} v^{\rho}\partial_{\rho} C_{\alpha\beta\gamma}(v_1,0,0,\dots)+\dots\,.
\label{Cexpan}
\ee
We are here interested in the three- and four-point correlation numbers. The
necessary results for the first two terms of \eqref{Cexpan} are presented
below. For a short representation, we introduce the function
$\mbox{\Large$\chi$}_{A,B}(x)=1$ if $x\in [A,B]$ and zero otherwise. In the
zeroth order, the structure constant in the $\bf{v}$ coordinates on the
solution of the string equation itself is
\begin{eqnarray}\label{strconstflat}
&C_{\alpha\beta\gamma}= \mbox{\Large$\chi$}_{1,q-1}(\alpha+\beta-\gamma)
\big(\!\!-\frac{v_1}{q}\big)^{\frac{\alpha+\beta+\gamma-q-1}{2}} \!\!\text{ if} \!\!\quad
\frac{\alpha+\beta+\gamma-q-1}{2}\in\mathbb{N},\!\text{ else 0}.
\end{eqnarray}
The first derivative $\partial_\rho C_{\alpha\beta\gamma}$
is given by
\begin{eqnarray}
\label{derivstrconst}
&\partial_\rho C_{\alpha\beta\gamma}=
\bigg[(q-\rho)\mbox{\Large$\chi$}_{1,\rho}(\alpha+\beta-\gamma)
+\frac{2q+\gamma-\alpha-\beta-\rho}{2}\mbox{\Large$\chi$}_{\rho+2,2q-\rho-2}(\alpha+\beta-\gamma)
\bigg]\times\nonumber\\
&\qquad\times\,\,\frac{2q-\alpha-\beta-\gamma-\rho}{2q}
\big(\!\!-\frac{v_1}{q}\big)^{\frac{\alpha+\beta+\gamma+\rho-2q-2}{2}}
\text{ if} \quad
\frac{\alpha+\beta+\gamma+\rho-2q-2}{2}\in\mathbb{N},\text{ else 0}.
\end{eqnarray}
In \eqref{strconstflat} and \eqref{derivstrconst}, $\mathbb{N}$ is the set
of nonnegative integers, and we assume the  ordering
$\rho\geq\alpha\geq\beta\geq\gamma$. Because both tensors are symmetric,
this information provides a complete answer. Some details of the derivation
are given in Appendix \bref{sec:StrConst}.

\section{Resonance transformations}
\label{sec:Reson}

In the continuous approach the integrated correlators of MLG are defined up to  
so-called contact terms, which are not determined by CFT methods \cite{Belavin:2008kv}.
On the other hand, any change of contact terms is equivalent to 
resonance transformations of the coupling parameters.
Explicitly, the resonance transformation has the form
\begin{equation}
t_{mn}=\lambda_{mn}+
\!\!\!\!\!\!\!\!\sum_{m_1,n_1}^{\delta_{mn}=\delta_{m_1n_1}+N}\!\!\!\!\!\!\!\!A^{(m_1n_1)}_N\mu^N \lambda_{m_1n_1}
+\!\!\!\!\!\!\!\!\!\sum_{m_1,n_1,m_2,n_2}^{\delta_{mn}=\delta_{m_1n_1}
+\delta_{m_2n_2}+N} \!\!\!\!\!\!\!\!A^{(m_1n_1,m_2n_2)}_N\mu^N \lambda_{m_1n_1}\lambda_{m_2n_2}+\dots\,.
\label{coupling}
\end{equation}
Here, $\mu$ and $\lambda_{mn}$ are the respective cosmological and Liouville
coupling constants, $N$ is a nonnegative integer, each pair $(m_i,n_i)$
satisfies\footnote{We always assume that this requirement is satisfied in
what follows.} $1\leq n_i\leq m_i\leq q-1$, and the constants
$A^{(ij)}_N,A^{(ij,kl)}_N,\dots$ are the parameters of the resonance
transformations. In \eqref{coupling}, the gravitational dimensions are given
by
\begin{equation}
\delta_{mn}=\frac{2q+1-|(q+1)m-q n|}{2q}\,,
\label{gravdim}
\end{equation}
and hence $\lambda_{mn}\sim\mu^{\delta_{mn}}$.

By regrouping the terms, we can write action \eqref{action ambigious} in the
form
\begin{equation}
S=S^{(0)}+\sum_{m,n} \lambda_{mn} S^{(mn)}+
\sum_{m_1,n_1,m_2,n_2} \lambda_{m_1n_1}\lambda_{m_2n_2} S^{(m_1n_1,m_2n_2)}+\dots\,.
\label{Slambda}
\end{equation}
In what follows, we call  the coefficients in expansion \eqref{Slambda}
{\it counterterms}.\footnote{We use this terminology from the
renormalization theory because the additional terms in the action play
exactly this role in responding to the shifts between the {\it bare}
parameters $t_{mn}$ and the {\it physical} parameters $\lambda_{mn}$.
According to this analogy, the selection rules for the correlators are regarded as
{\it renormalization conditions}.}

All required information concerning correlation functions is encoded in
generating function \eqref{Z}. Namely, the correlation numbers are related
to the coefficients in the coupling constant decomposition of the
generating function
\begin{equation}
Z=Z_0+\sum_{m_1,n_1} \lambda_{m_1n_1} Z_{m_1n_1}
+\sum_{m_1,n_1,m_2,n_2} \lambda_{m_1n_1}\lambda_{m_2n_2} Z_{m_1n_1,m_2,n_2}+\dots\,.
\end{equation}
In what follows, we often use the short notation $\lambda_{m_i,n_i}=\lambda_{i}$ and $Z_{m_in_i,m_jn_j,...}=Z_{ij...}$.

\subsection{First-order counterterms}
\lem{Lemma4} The first-order counterterms are given by
\begin{eqnarray}
S^{(m_1n_1)}=\underset{y=\infty}{\text{res}} \sum_{N=0}^{\lfloor  \frac{m-n}{2}\rfloor} A^{(m_1n_1)}_N
\mu^N Q^{\frac{(q+1)m-q (n+2N)}{q}}(y)\,,
\label{countpart1}
\end{eqnarray}
where $m=m_1$,  $n=n_1$, and $A^{(m_1n_1)}_N$ are the coefficients in the
resonance relations of the coupling constants ($A^{(m_1n_1)}_0=1$).
\elem
This statement follows trivially from \eqref{gravdim}.

\subsection{Second-order counterterms}

Taking \eqref{coupling} and \eqref{Slambda} into account, we can derive the
explicit form of $S^{(m_1n_1,m_2n_2)}$ from definition \eqref{action ambigious}.
The conditions on the pairs $(m_1,n_1)$ and $(m_2,n_2)$ for which there
exist some $(m,n)$ in the Kac table of $M_{q,q+1}$ such the corresponding
three gravitational dimensions are subject to the resonance balance are
formulated below.

\lem{Lemma5} The second-order counterterms are given by
\begin{eqnarray}
S^{(m_1n_1,m_2n_2)}=\underset{y=\infty}{\text{res}} \sum_{N=0}^{\lfloor  \frac{m-n}{2}\rfloor}
A^{(m_1n_1,m_2n_2)}_N
\mu^N Q^{\frac{(q+1)m-q (n+2N)}{q}}(y)\,,
\label{countpart2}
\end{eqnarray}
where
\begin{align}
&m=m_1+m_2-1, \qquad &n=n_1+n_2+1 \qquad \text{for}\qquad m_1+m_2\leq q\,,\\
&m=m_1+m_2-q-1, \qquad &n=n_1+n_2-q \qquad \text{for}\qquad m_1+m_2> q\,.
\label{Lemma5}
\end{align}
\elem

\prf
We seek solutions of the dimensional balance requirement
\begin{equation}
\delta_{m_1n_1}+\delta_{m_2n_2} + N =\delta_{mn}\,,
\end{equation}
where $N$ is a nonnegative integer and $n\leq m\leq q-1$. It is convenient
to set $d=m-n$ and $d_i=m_i-n_i$. Explicitly, this requirement gives
\begin{equation}
d_1+d_2+\frac{m_1+m_2}{q}=d+2N+2+\frac{m+1}{q}\,.
\end{equation}
\begin{enumerate}
\item
Let $m_1+m_2<q$. If $m+1=q$, then there is no solution because the
noninteger factor $\frac{m_1+m_2}{q}$ cannot be compensated.  If $m+1<q$,
then
\begin{eqnarray}
\begin{cases}
d_1+d_2=d+2N+2,\\
m_1+m_2=m+1.
\end{cases}
\end{eqnarray}
This gives $m=m_1+m_2-1$ and $n=n_1+n_2+1+2N$.
\item
Let $m_1+m_2=q$. If $m+1<q$, then there is no solution because the
noninteger factor $\frac{m+1}{q}$ cannot be compensated. If $m+1=q$, then
\begin{eqnarray}
\begin{cases}
d_1+d_2+1=d+2N+3,\\
m_1+m_2-q=0.
\end{cases}
\end{eqnarray}
This gives again $m=m_1+m_2-1$ and $n=n_1+n_2+1+2N$, where $m_1+m_2=q$, and
this case can hence be joined with the preceding case.
\item
Let $q< m_1+m_2\leq 2q-2$. If $m+1=q$, then there is no solution because the
noninteger factor $\frac{m_1+m_2-q}{q}$ cannot be compensated. If $m+1<q$,
then
\begin{eqnarray}
\begin{cases}
d_1+d_2+1=d+2N+2,\\
m_1+m_2-q=m+1,
\end{cases}
\end{eqnarray}
This gives $m=m_1+m_2-q-1$ and $n=n_1+n_2-q+2N$.
\end{enumerate}
\epr

\section{Partition function and one-point correlators}
\label{sec:onepoint}

There is much evidence that the approach based on the Douglas string
equation formulated in the preceding section provides an alternative
description of MLG. A general proof of this statement is not yet available,
but this conjecture can be checked by comparison with the results of
computing directly in the framework of the continuous approach. Below, we
provide further support of the hypothesis by performing some checks on the
level of three- and four-point correlators. The last case is  most important
because the continuous approach here first requires nontrivial  integration
over the moduli space \cite{BelAlZam}.

For convenience, we start our analysis of the correlation functions by
presenting the results for the zero-, one-, and two-point correlators
found in \cite{VBelavin:2014fs}. The consideration in this and in the
following sections is based on the following statement.

\lem{Lemma3}
On the line $v_{i>1}=0$,
\begin{eqnarray}
\begin{cases}
&k-\text{even}:\qquad
\frac{\partial \theta_{\lambda,k}}{\partial v_{\alpha}}=\delta_{\lambda,\alpha} \,
x_{\lambda,k}\,\big(-\frac{v_1}{q}\big)^{\frac{k}{2}q},
\\
&k-\text{odd}:\qquad
\frac{\partial \theta_{\lambda,k}}{\partial v_{\alpha}}=\delta_{\lambda,q-\alpha} \,
y_{\lambda,k}\,\big(-\frac{v_1}{q}\big)^{\frac{k-1}{2}q+\lambda},
\end{cases}
\label{Lemma3}
\end{eqnarray}
where
\begin{eqnarray}\label{xy}
x_{\lambda,k}=\frac{1}{\big(\frac{\lambda}{q}\big)_{\frac{k}{2}}
\big(\frac{k}{2}\big)!}
\qquad\text{and}\qquad
y_{\lambda,k}=-\frac{1}{\big(\frac{\lambda}{q}\big)_{\frac{k+1}{2}}
\big(\frac{k-1}{2}\big)!} \,.
\label{Lemma31}
\end{eqnarray}
\elem
In \cite{VBelavin:2014fs}, this result was derived from recurrence
relations \eqref{Recursion0}.

\subsection{Partition function}
To define normalization independent quantities (or invariant cross ratios),
we need the explicit form of the partition function
\begin{equation}
Z_{0}=\frac{1}{2}\int_0^{{\bf v}_*^{(0)}} d v^{\gamma} C_{\gamma}^{\alpha\beta} \frac{\partial S^{(0)}}{\partial v^\alpha}
\frac{\partial S^{(0)}}{\partial v^\beta}\,,
\label{Z0}
\end{equation}
where ${\bf v}_*^{(0)}$ denotes the zeroth-order term of the expansion in
coupling constants of the solution of the string equation. 

To find it, we
write the zeroth-order term of the expansion of the action explicitly in
terms of deformed flat coordinates \eqref{theta},
\begin{equation}\label{S0}
S^{(0)}=\underset{y=\infty}{\text{res}}\bigg[Q^{\frac{2q+1}{q}}+\mu Q^{\frac{1}{q}}\bigg]=
-\frac{ \theta_{1,2}}{c_{1,2}}-\mu\frac{\theta_{1,0}}{c_{1,0}}\,.
\end{equation}
It can be seen from $\eqref{Lemma3}$ that equations \eqref{streqflat} for
$\alpha<q-1$ are solved automatically by the ansatz
${\bf v}_*=(v_{*1},0,0,\dots)$, and we are hence left with the equation
\begin{equation}
\frac{\partial S^{(0)}}{\partial v_1}=-
\frac{1}{c_{1,2}} \frac{\partial \theta_{1,2}}{\partial v_1}-
\frac{\mu}{c_{1,0}} =0\,,
\label{streq0}
\end{equation}
which gives
\begin{equation}
\mu = \frac{(1+q)(1+2q)}{q} \bigg(\!\!- \frac{v_{*1}}{q}\bigg)^q\,.
\end{equation}
This equation defines the zeroth-order term in the expansion of the
appropriate solution of the string equation. The integration contour in
\eqref{Z0} can be taken along the axis $v_1$ because on the line
$\frac{\partial S^{(0)}}{\partial v_k}=0$  for $k>1$, we are left with the
term containing  $C_{q-1}^{q-1,q-1}=1$. An explicit calculation gives
\begin{equation}
Z_{0}=  \frac{(1+q)(1+2q)}{q^{2q+2}} \, v_{*1}^{2q+1}\,.
\label{Z0result}
\end{equation}

\subsection{One-point functions}
For the general one-point correlators, we obtain
\begin{equation}
Z_{mn}=\int_0^{{\bf v}_*^{(0)}} d v^{\gamma} C_{\gamma}^{\alpha\beta} \frac{\partial S^{(0)}}{\partial v^\alpha}
\frac{\partial S^{(mn)}}{\partial v^\beta}\,.
\label{Zmn}
\end{equation}
Taking into account that $C_{\alpha}^{q-1,\beta}=\delta_{\alpha\beta}$ on
the line $v_{k>1}=0$ and using the results in the preceding subsection, we
more explicitly obtain
\begin{eqnarray}
Z_{mn}
=\int_0^{{\bf v}_*^{(0)}} C_{q-1}^{q-1,\gamma} \frac{\partial S^{(0)}}{\partial v^{q-1}}
\frac{\partial S^{(mn)}}{\partial v^{\gamma}} d v_{1}=
\int_0^{{\bf v}_*^{(0)}}\frac{\partial S^{(0)}}{\partial v_{1}}
\frac{\partial S^{(mn)}}{\partial v_{1}} d v_{1}.
\label{Zmn1}
\end{eqnarray}
Based on \refl{Lemma3}, we can conclude \cite{VBelavin:2014fs} that the
one-point correlation numbers are equal to zero for all
fields.\footnote{This statement holds up to usual indefiniteness related to
the correlators having integer gravitational dimensions \cite{Belavin:2008kv}.}

\section{Two-point correlators}
\label{sec:twopoint}
An essential consistency requirement is that the two-point correlators be
diagonal. This allows defining the first-order counterterms in the resonance
transformations. Differentiating \eqref{Z} twice, we obtain
\begin{equation}
Z_{m_1n_1,m_2n_2}= \int_0^{{\bf v}_*^{(0)}} d v_1 C_{q-1}^{\alpha\beta}
\frac{\partial S^{(m_1n_1)}}{\partial v^{\alpha}} \frac{\partial S^{(m_2n_2)}}{\partial v^{\beta}}+
\int_0^{{\bf v}_*^{(0)}} d v_1 C_{q-1}^{\alpha\beta}\frac{\partial S^{(0)}}{\partial v^{\alpha}}
\frac{\partial S^{(m_1n_1,m_2n_2)}}{\partial v^{\beta}}.
\end{equation}
The second term can be nonzero only for the correlators that have integer
gravitational dimensions, which we do not consider. The first term gives
\begin{equation}
Z_{m_1n_1,m_2n_2}=\sum_{\gamma=1}^{q-1} (-q)^{1-\gamma}
\int_0^{{\bf v}_*^{(0)}}  \!\!\!\!\!d v_1\, v_1^{\gamma-1} \,
\frac{\partial S^{(m_1n_1)}}{\partial v_{\gamma}}\, \frac{\partial S^{(m_2n_2)}}{\partial v_{\gamma}}.
\end{equation}
It was found in \cite{VBelavin:2014fs} that for $v_{i>1}=0$,
\begin{eqnarray}\label{Smn_alpha}
\frac{\partial S^{(mn)}}{\partial v_{\alpha}}(v_1)=\begin{cases}
\delta_{m,\alpha} v_{*1}^{\frac{m-n}{2}q} (-q)^{\frac{\alpha-1}{2}}N_{mn} P_{\frac{m-n}{2}}^{(0,\frac{m-q}{q})}(t),&\!\!\!(m-n)\text{ even},\\
\delta_{m,q-\alpha} v_{*1}^{\frac{m-n-1}{2}q+m} (-q)^{\frac{\alpha-1}{2}}N_{mn} \big(\frac{1+t}{2}\big)^{\frac{m}{q}}
P_{\frac{m-n-1}{2}}^{(0,\frac{m}{q})}(t),&\!\!\!(m-n)\text{ odd},
\end{cases}
\label{Smn}
\end{eqnarray}
where the new variable
\begin{eqnarray}
\label{t}
t=2\bigg(\frac{v_1}{v_{*1}}\bigg)^q-1\,,
\label{tv}
\end{eqnarray}
$P_n^{(0,b)}(t)$ are the Jacobi polynomials (see Appendix \bref{sec:Jacobi}),
and $N_{mn}$ denotes $t$-independent factors. Its explicit form is not
relevant for our further consideration. With this result, the diagonality
condition for the two-point correlators,
\be
Z_{m_1n_1,m_2n_2}\sim\delta_{m_1,m_2}\delta_{n_1,n_2}\,,
\ee
becomes equivalent to the orthogonality condition for the Jacobi polynomials.
Calculating the diagonal two-point functions is straightforward,
\begin{eqnarray}
\qquad Z_{mn,mn}=\frac{N_{mn}^2}{(m- n)q+m}\, v_{*1}^{(m- n)q+m}\,.
\label{Z12}
\end{eqnarray}

\section{Three-point correlators}
\label{sec:threepoint}

One important change when we proceed to the level of three-point functions
is that the derivative of the upper integration limit should be taken into
account. Also for the first time, the contribution of the second-order
counterterms arise on the level of three-point correlation functions. Using
the same arguments based on the string equation and nonanalyticity
requirements, we obtain the expression
\begin{eqnarray} \label{Z123_0}
Z_{123} =\sum_{\sigma}\int_0^{{\bf v}_*^{(0)}} \!\!\!\!\!\!
d v^{\gamma} C_{\gamma}^{\alpha\beta} \frac{\partial S^{(\sigma(1))}}{\partial v^{\alpha}}
\frac{\partial S^{(\sigma(2)\sigma(3))}}{\partial v^{\beta}}+
C_{\gamma}^{\alpha\beta} \frac{\partial {\bf v}_{*}^{\gamma}}{\partial \lambda_3}
\frac{\partial S^{(1)}}{\partial v^{\alpha}}\frac{\partial S^{(2)}}{\partial v^{\beta}},
\label{Z123start}
\end{eqnarray}
where we replace the indices $(m_i,n_i)$ with index $i$ and the sum ranges
permutations of the set $\{1,2,3\}$. In \eqref{Z123start} and below, we
always assume that the nonintegral part is evaluated on the solution of the
string equation for all couplings equal to zero. In the rest of this paper,
we use  Latin indices exclusively in this sense, and the index zero means
the zeroth-order term in the coupling constants expansion. The other terms
in \eqref{Z123} disappear because they contain $\frac{\partial S^{(0)}}
{\partial v^{\alpha}}(v_*)=0$. It follows from the string equation that
\begin{equation}
\frac{\partial S^{(i)}}{\partial v^{\alpha}}+
\frac{\partial^2 S^{(0)}}{\partial v^{\alpha}\partial v^{\gamma}} \,
\frac{\partial v_*^{\gamma}}{\partial \lambda_i } =0\,,
\end{equation}
which gives
\begin{equation}\label{partialv}
\frac{\partial v_*^{\gamma}}{\partial \lambda_{i} } =
T^{\gamma\beta} \frac{\partial S^{(i)}}{\partial v^{\beta}}\,.
\end{equation}
The inverse matrix ($T^{\alpha\gamma}M_{\gamma\beta}=\delta^{\alpha}_{\beta}$)
\begin{equation}
M_{\alpha\beta}=- \frac{\partial^2 S^{(0)}}{\partial v^{\alpha}\partial v^{\beta}}\,,
\end{equation}
can be calculated using \eqref{S0}. The second term does not contribute to
$M_{\alpha\beta}$, because $\frac{\partial^2\theta_{1,0}}{\partial v^{\alpha}
\partial v^{\beta}}=\frac{\partial^2 v_1}{\partial v^{\alpha}
\partial v^{\beta}}=0$. For the first term, we use \eqref{Lemma3},
\begin{eqnarray}
M_{\alpha\beta}=\frac{1}{c_{1,2}} \frac{\partial^2 \theta_{1,2}}{\partial v^{\alpha}\partial v^{\beta}}=
\frac{1}{c_{1,2}}\, v_1 C^1_{\alpha\beta}\,.
\end{eqnarray}
With this result, we can easily find the inverse matrix $T^{\gamma\beta}$:
\be
T^{\alpha\beta}=\frac{c_{1,2}}{v_1}\bigg(\!\!\!-\frac{v_1}{q}\bigg)^{1-\alpha}\delta_{\alpha,\beta}\,.
\ee
The three-point function becomes
\begin{eqnarray}
Z_{123} =
\sum_{\sigma}\int_0^{{\bf v}_*^{(0)}} d v^{\gamma} C_{\gamma}^{\alpha\beta} \frac{\partial S^{(\sigma(1))}}{\partial v^{\alpha}}
\frac{\partial S^{(\sigma(2)\sigma(3))}}{\partial v^{\beta}}+
C_{\rho}^{\alpha\beta} T^{\rho\gamma}
\frac{\partial S^{(1)}}{\partial v^{\alpha}}\frac{\partial S^{(2)}}{\partial v^{\beta}}
\frac{\partial S^{(3)}}{\partial v^{\gamma}}\,,
\label{3point1}
\end{eqnarray}
where the second term is evaluated on the solution of the string equation.

\subsection{Fusion rules and three-point functions}
We first formulate a useful consequence of the fusion rules for the
three-point function in unitary minimal models $M_{q,q+1}$. We recall that
the primary fields $\Phi_{mn}$  are labeled by $m=1,\dots,q-1$ and $n\leq m$.
For the three point function
\be
G=\langle \Phi_{m_1n_1}\Phi_{m_2n_2}\Phi_{m_3n_3}\rangle\,,
\label{threepoint}
\ee
the fusion rules are satisfied, i.e., $G\neq 0$ if  $P=(m_1n_1,m_2n_2,m_3n_3)$ or one of its
reflection images generated by $(m_i,n_i)\to(q-m_i,q+1-n_i)$  belongs to the
region
\be
F =\{(m_1n_1,m_2n_2,m_3n_3)\}
\ee
such that for some permutation $(i,j,k)$ of the set $\{1,2,3\}$
\be
\begin{cases}
m_k \in [|m_i-m_j|+1\,:2:\,\text{min}(m_i+m_j-1,2q-1-m_i-m_j)]\,,\\
\,\,\, n_k \in\, [|n_i-n_j|+1\,:2:\,\text{min}(n_i+n_j-1,2q+1-n_i-n_j)]\,,
\end{cases}
\label{fusionregion}
\ee
where $:2:$ denotes step two. We can classify different cases with respect
to the parities of $\sum_{i=1}^3m_i$ and $\sum_{i=1}^3n_i$. We can easily
see the following consequence of the fusion rules for the unitary minimal
models.
\prop{Conseq}
In the case where both $\sum_im_i$ and $\sum_in_i$ are even, the three-point
function  $G=0$.
\eprop
Indeed, for any choice of the reflection images of the fields, one of the
parities $\sum_im_i$ or $\sum_in_i$ is even, which is forbidden
by \eqref{fusionregion}.

\subsection{Comparing with MLG results}
\label{sec:genpatt}

For the physically relevant nonanalytic correlators, there are two
possibilities:
\begin{enumerate}
\item
The fusion rules are not satisfied, the case {\it nonphysical} below. In
this case, \eqref{3point1} must give zero whenever the resonance
transformations permit it. This requirement allows defining the second-order
counterterms $S^{(12)}$.
\item
The parameters $\{m_i,n_i\}$ of the three-point function $Z_{123}$ satisfy
the fusion rules described above. We call this the {\it physical region}. In
this case, \eqref{3point1} must give a result the same as the result derived
in the continuous approach \cite{BelAlZam}. This can be achieved if the two
conditions are satisfied:
\begin{itemize}
\item[(a)]
The integral part of \eqref{3point1} in this region is zero.
\item[(b)]
The nonintegral part of \eqref{3point1} is nonzero and leads to the correct
answer for the universal ratios.
\end{itemize}
\end{enumerate}

Without loss of generality (interchanging the pairs if necessary), we can
fix $2\leq m_1\leq m_2 \leq m_3$. The general analysis of the three-point
sector requires considering four different domains: all three fields are
even, one field is odd, two fields are odd, all three fields are odd. Below,
we analyze $(1)$ and $(2)$ in detail in the  domain where all fields are
even. Moreover, we impose the additional constraint
\be
m_{12}\leq m_{13} \leq m_{23}\leq q.
\label{region}
\ee
We note that the three-point functions are always nonanalytic in this
domain. Indeed, the dimension
\begin{equation}
[Z_{123}]=[Z]-\sum_{i=1}^3\delta_i=-1+\frac{m_1+m_2+m_3-1}{2q}+\sum_{i=1}^3 \frac{m_i-n_i}{2}\,,
\label{dimZ123}
\end{equation}
where $m_1+m_2+m_3\leq 2q-1$ and hence \eqref{dimZ123} is not integer.

\subsubsection{Nonphysical region}
Using the explicit form of the structure constants and the properties of
first- and second-order counterterms \eqref{Smn} and \eqref{Lemma5}, we
write the integral part IP of \eqref{3point1}:
\begin{equation}
\text{IP}=\sum_{\sigma}\sum_{\gamma=1}^{q-1} \delta_{m_i,\gamma}\delta_{m_j+m_k-1,\gamma}
\int_0^{{\bf v}_{*}^{(0)}} d v_1 \big(\!\!-\frac{v_1}{q}\big)^{\gamma-1}
\frac{\partial S^{(\sigma(i))}}{\partial v_{\gamma}} \frac{\partial S^{(\sigma(j),\sigma(k))}}{\partial v_{\gamma}}(v_1) \,.
\end{equation}
To find the second-order counterterm, we take
\begin{equation}
m_3=m_1+m_2-1\,.
\end{equation}
It can be seen that the terms with permutations disappear. Before we use
change \eqref{tv}, it is convenient to express the second-order
counterterms also in terms of the dimensionless functions
$X^{(m_1n_1,m_2n_2)}(t)$:
\begin{equation}
\frac{\partial S^{(m_1n_1,m_2n_2)}}{\partial v_{m_3}}(v_1)=N_1 N_2
\,c_{1,2} 2^{-\frac{m_3}{2}}
(-q)^{\frac{2q-3+m_3}{2}}
 \,X^{(m_1n_1,m_2n_2)}(t)v_{*1}^{\frac{m_3-n_1-n_2-1}{2}q}\,\,.
\end{equation}
The explicit form of the nonintegral part NIP in \eqref{3point1} is
\begin{align}
\text{NIP}&=  N_1 N_2 N_3 \frac{q^{q+1}}{(1+q)(1+2q)} v_{1*}^{\sum_i (\frac{m_i-n_i}{2}q+\frac{m_i}{2})-\frac{1+2q}{2}}\,.
\label{Z123}
\end{align}
Combining the integral and nonintegral parts, we obtain
\begin{equation}
\begin{aligned}
Z_{123} =N_1 N_2 N_3\, c_{1,2} (-q)^{q-2}
\bigg[\int_{-1}^{1} d t
(1+t)^{\frac{m_3-q}{q}}
P_{\frac{m_3-n_3}{2}}^{(0,\frac{m_3-q}{q})}(t) \,  X^{(m_1n_1,m_2n_2)}(t)-1 \bigg]\,.
\end{aligned}
\label{Xcond}
\end{equation}
The degree of this polynomial  $X^{(m_1n_1,m_2n_2)}(t)$ is
\begin{equation}
\text{deg}\,X^{(m_1n_1,m_2n_2)}=\frac{m_1-n_1+m_2-n_2-2}{2}\,.
\label{degX}
\end{equation}
Because $m_3-n_3$ is even,  $n_3-n_1-n_2$ should be odd:
\begin{equation}
n_3=n_1+n_2+1+ 2s\,,\qquad  s\in \mathbb{Z}\,.
\end{equation}

We consider the region $n_3\geq n_1+n_2$, where the three-point function
should be zero according to the fusion rules. Because of the parity
requirement, $n_3$ takes values from $n_1+n_2+1$ to $m_3$ with step $2$ in
this region. The degree of the Jacobi polynomial in \eqref{Xcond} then
changes from $0$ to $\text{deg}\,X^{(12)}$ with step $1$. Taking the
completeness property of the Jacobi polynomials into account, we obtain
\begin{equation}
X^{(12)}(t)=
\sum_{k=0}^{\text{deg}\, X^{(12)}} \!\!\bigg(q k +\frac{m_1+m_2-1}{2}\bigg)\, P_k^{(0,\frac{m_1+m_2-1}{q}-1)}(t)\,,
\label{X1}
\end{equation}
where the coefficients are fixed (using orthogonality \eqref{JacobiScalar})
from the requirement $Z_{123}=0$. Equation \eqref{JacobiProp1} can be used
to obtain the representation\footnote{We note that according to the results
in \refl{Lemma5}, $m_1+m_2-n_1-n_2\geq 2$.}
\begin{equation}
X^{(12)}(t)= q
(1+t)^{1-\frac{m_1+m_2-1}{q}}\frac{d}{d t}\bigg[(1+t)^{\frac{m_1+m_2-1}{q}}
 P_{\frac{m_1-n_1+m_2-n_2}{2}-1}^{(0,\frac{m_1+m_2-1}{q})}(t)\bigg]\,.
\label{X3}
\end{equation}

\subsubsection{Physical region}
We must first ensure that if the fusion rules are satisfied, then the
nonintegral part is nonzero. From \eqref{strconstflat}, we find that
$\text{NIP}\neq 0$ if
\begin{align}
&m_3=m_1+m_2-1-2 s\,, \qquad s=0,1,2,\dots\,,
\label{cond0}
\end{align}
and
\begin{align}
\label{cond1} &1\leq q-m_3+m_2-m_1\leq q-1\,.
\end{align}
Taking \eqref{region} into account, we note that condition \eqref{cond0} is
equivalent to $m_3\leq \min(m_1+m_2-1,2q-1-m_1-m_2)$. The right condition
in \eqref{cond1} gives $m_2-m_1+1\leq m_3$, while the left condition is
satisfied because $q-m_3+m_2-m_1\geq q-m_3$. Hence, if the fusion rules are
satisfied, then the nonintegral part in \eqref{3point1} is nonzero.

The second step is to verify that the integral part in \eqref{3point1} is
absent when the fusion rules are satisfied. It follows from \refl{Lemma5}
that $\text{IP}\!=0$ if $m_3\neq m_1+m_2-1$. Hence, all we need to verify
is the case $m_3=m_1+m_2-1$. From the fusion rules, we have
\be
|n_1-n_2|+1 \leq n_3 \leq n_1+n_2-1\,,
\ee
where the choice of the right-hand side takes $n_i\leq m_i$ and
\eqref{region} into account. From \eqref{degX}, we derive
\be
\text{deg}\,  X^{(m_1n_1,m_2n_2)}(t) < \text{deg}\, P_{\frac{m_3-n_3}{2}}^{(0,\frac{m_3-q}{q})}(t)\,,
\ee
and using the completeness and orthogonality properties of Jacobi
polynomials, we conclude that the integral part is equal to zero.

The final step is to check the result for three-point universal ratio. For
general $(q,p)$  \cite{BelAlZam}, it has the form
\begin{equation}
\frac{\langle \langle O_1 O_2 O_3\rangle\rangle^2}{\prod_{i=1}^3
\langle \langle O_i \rangle\rangle}
=\frac{\prod_{i=1}^3|p m_i-q n_i|}{p(p+q)(p-q)}\,,
\end{equation}
where $\langle\langle\dots\rangle\rangle=\frac{\langle\dots\rangle}{\langle 1 \rangle}$.
For $p=q+1$, this expression coincides with the three-point universal ratio
obtained in the dual approach
\begin{equation}
\frac{(Z_{123})^2 Z_0}{Z_{11}Z_{22}Z_{33}}=\frac{\prod_{k=1}^3\big((q+1)m_i-q n_i\big)}{(1+q)(1+2q)}\,.
\end{equation}
Here, we use \eqref{Z123}, \eqref{Z12}, and \eqref{Z0result}.

\section{Four-point correlators}
\label{sec:fourpoint}

In this section, we perform some checks for the four-point correlators. We
are mainly focused on the following puzzle. From the preceding consideration,
it might appear that the role of the parameters $n_i$ is somewhat suppressed
with respect to the role of the parameters $m_i$. Indeed, apart from the
normalization, it seems that all they do is choose the region according to
the parities $m_i-n_i$, and the values of $n_i$ seem irrelevant. Below, we
demonstrate how the balance between $m_i$ and $n_i$ is recovered on the
four-point level.

The general expression for the four-point correlator is
\be
Z_{1234}=Z_{1234}^{\text{NIP}}+Z_{1234}^{\text{IP}}\,,
\label{4pointgeneral}
\ee
where\footnote{To make our formulas less cumbersome, we just remember that
the resulting expression is to be calculated on the solution of the string
equation. Therefore, in particular, there is no term containing
$S^{(0)}_\alpha$ in the nonintegral part.}
\begin{eqnarray}
Z_{1234}^{\text{NIP}}=
\frac{\partial^2 v^\gamma_*}{\partial \lambda_3\partial \lambda_4}
C_{\gamma}^{\alpha\beta} S^{(2)}_\alpha S^{(1)}_\beta+\nonumber
\frac{\partial v^\gamma_*}{\partial \lambda_3}\frac{\partial v^\delta_*}{\partial \lambda_4} C_\gamma^{\alpha\beta}
\big(S_\beta^{(1)} S_{\alpha\delta}^{(2)}+ S_{\beta\delta}^{(1)} S_\alpha^{(2)} \big)+\nonumber\\
\qquad\qquad\qquad+\frac{\partial v^\gamma_*}{\partial \lambda_3} \frac{\partial C_\gamma^{\alpha\beta}}{\partial \lambda_4}
S_\alpha^{(2)}  S_\beta^{(1)} +\frac{\partial v^\gamma_*}{\partial \lambda_4}
C_{\gamma}^{\alpha\beta}
\big( S^{(12)}_\alpha S^{(3)}_\beta+\text{permutations}\big)\,,
\label{4pointNIP}
\end{eqnarray}
and
\be
Z_{1234}^{\text{IP}}=\frac{1}{2}\int_0^{v_{*1}} \,d v^\gamma C_\gamma^{\alpha\beta}
\big(S_\alpha^{(1234)}S_\beta^{(0)}+ S_\alpha^{(123)}S_\beta^{(4)}+S_\alpha^{(12)}S_\beta^{(34)}+\text{permutations}\big)\,.
\label{4pointIP}
\ee
In \eqref{4pointNIP} and \eqref{4pointIP}, we use $S_{\alpha}^{(i\dots)}=
\frac{\partial S^{(i\dots)}}{\partial v^\alpha}$ and
$S_{\alpha\beta}^{(i\dots)}=\frac{\partial^2 S^{(i\dots)}}
{\partial v^\alpha\partial v^\beta}$. According to the general pattern
sketched in \bref{sec:genpatt}, we assume that the integral part
$Z_{1234}^{\text{IP}}=0$ in the region where the fusion rules are satisfied.

We consider the case where there are no higher-order counterterms starting
from the second order and the first derivatives of the structure constant
are zero,\footnote{This requirement gives, of course, additional restrictions on the
parameters $m_i,n_i$ of the four-point correlator.} and the term with
\be
\frac{\partial C_\gamma^{\alpha\beta}}{\partial \lambda_i}=
T^{\rho\eta}\partial_\rho C_{\gamma}^{\alpha\beta} S^i_\eta
\ee
is hence absent. Under these assumptions, only the first two terms in
\eqref{4pointNIP} survive:
\be
Z_{1234}=\frac{\partial^2 v_*^\gamma}{\partial \lambda_3\partial \lambda_4}
C_{\gamma}^{\alpha\beta} S^{(2)}_\alpha S^{(1)}_\beta+
\frac{\partial v_*^\gamma}{\partial \lambda_3}
\frac{\partial v_*^\delta}{\partial \lambda_4} C_\gamma^{\alpha\beta}
\big(S_\beta^{(1)} S_{\alpha\delta}^{(2)}+ S_{\beta\delta}^{(1)} S_\alpha^{(2)} \big) \,.
\ee
Similar to \eqref{partialv}, we can find second derivatives of the solution
of the string equation
\be
\frac{\partial^2 v^\gamma_*}{\partial \lambda_i \partial \lambda_j}=
T^{\gamma\rho}T^{\sigma\chi}\big(
S_{\rho\sigma}^{(i)} S_\chi^{(j)}+ S_{\rho\sigma}^{(j)} S_{\chi}^{(i)}
+T^{\eta\delta}S^{(0)}_{\rho\sigma\eta} S_{\chi}^{(i)} S_{\delta}^{(j)}
\big)\,.
\ee
Combining all together, we obtain the structure of the four-point
correlator:
\begin{eqnarray}\label{fourpoint1}
Z_{1234}=&
 T^{\gamma\rho}T^{\sigma\chi} C_{\gamma}^{\alpha\beta} S^{(1)}_\alpha S^{(2)}_\beta \big(
S_{\rho\sigma}^{(3)} S_\chi^{(4)}+ S_{\chi}^{(3)} S_{\rho\sigma}^{(4)}\big)+\nonumber\\
 &T^{\gamma\mu}T^{\delta\nu} C_\gamma^{\alpha\beta}
\big(S_\beta^{(1)} S_{\alpha\delta}^{(2)}+ S_{\beta\delta}^{(1)} S_\alpha^{(2)} \big)S_{\mu}^{(3)} S_\nu^{(4)}
+\\
& T^{\gamma\rho}T^{\sigma\chi}T^{\eta\delta}C_{\gamma}^{\alpha\beta}
S^{(0)}_{\rho\sigma\eta} S^{(1)}_\alpha S^{(2)}_\beta S_{\chi}^{(3)} S_{\delta}^{(4)}\,.\nonumber
\end{eqnarray}
Even without the contribution of the higher counterterms, we see a few new
objects in this expression that require additional calculations. The details
of the calculations are in Appendix \bref{sec:App4point}. For the third
derivative $S^{(0)}_{\rho\sigma\eta}$, we obtain
\be
S^{(0)}_{\beta\rho\eta}=\frac{q-\beta-\rho-\eta-1}{2 c_{1,2}} C_{\beta\rho\eta}\,,
\label{thirdderivS0}
\ee
and the second derivative $S^{(i)}_{\alpha\beta}$ is given by
\be
S^{(mn)}_{\alpha\beta}=\frac{1}{2} C_{\alpha\beta}^{m}
\bigg(\!\!-\frac{v_1}{q}\bigg)^{\!\!m+1-q}\!\! R_{mn}\,,
\label{2derivSi}
\ee
where there is no summation over $m$ and
\be
R_{mn}=\frac{1}{2} (m - n) (2 m + q (m - n ))\,.
\ee
In the calculation, we find the three basic structures
\begin{align}\label{F1}
&F_1(\chi,\xi,\mu,\nu)=\sum_{\gamma=1}^{q-1} \theta(\gamma,\chi,\xi)
\theta(\gamma,\mu,\nu)\,,\\
&F_2(\chi,\xi,\mu,\nu)=\sum_{\gamma=1}^{q-1} \theta(\gamma,\chi,\xi)
\theta(q\!-\!\gamma,\mu,\nu)\label{F2}\,,\\
&F_3(\chi,\xi,\mu,\nu)=\sum_{\gamma=1}^{q-1} \gamma\,\theta(\gamma,\chi,\xi)
\theta(\gamma,\mu,\nu)\label{F3}\,.
\end{align}
Here, each $\theta(\alpha,\beta,\gamma)$ is related to one of the structure
constants in \eqref{fourpoint1}, \eqref{thirdderivS0}, or \eqref{2derivSi}.
We recall that it is explicitly defined as a symmetric tensor such that
$\theta(\alpha,\beta,\gamma)=\mbox{\Large$\chi$}_{1,q-1}(\alpha+\beta-\gamma)$
if  $\alpha\geq\beta\geq\gamma$. After some computations (the details can be
found in Appendix \bref{sec:App4point}), we obtain
\begin{align}\label{F1_result}
&F_1(\chi,\xi,\mu,\nu)\!=\!\\
&q-\frac{\!\xi\! +\!\chi\!+\!\mu\! +\!\nu }{4}
-\frac{|\chi\! -\!\xi |+|\mu\! -\!\nu |+
\big{|}|\chi\! -\!\xi|-|\mu\! -\!\nu |\big{|}+|\mu \!+\!\nu \!-\!\xi\! -\!\chi |}{4}\,,\nonumber\\
&F_2(\chi,\xi,\mu,\nu)\!=\!\label{F2_result}\\
&q-\frac{\!\xi\! +\!\chi\!+\!\mu\! +\!\nu}{4}
-\frac{|\chi\! -\!\xi |+|\mu\! -\!\nu |+
\big{|}q\!+\!|\chi\! -\!\xi|\!-\!\mu\! -\!\nu \big{|}+
\big{|}q\!+\!|\mu\! -\!\nu|\!-\!\chi\! -\!\xi \big{|}}{4}\,,\nonumber\\
&F_3(\chi,\xi,\mu,\nu)\!=F_1(\chi,\xi,\mu,\nu)\times\label{F3_result}\\
&
\!\!\times\!\bigg[q+\frac{\chi\!+\!\xi\!+\!\mu\! +\!\nu\!}{4}-\frac{|\chi\! -\!\xi |+|\mu\! -\!\nu |+
\big{|}|\chi\! -\!\xi|-|\mu\! -\!\nu |\big{|}-|\mu \!+\!\nu \!-\!\xi\! -\!\chi |}{4}-1
\bigg]\,.
\nonumber
\end{align}
In terms of these functions with the convention $R_i=R_{m_in_i}$, the
four-point correlator is
\begin{eqnarray}\label{fourpoint2}
&Z_{1234}=N\bigg[R_{1} F_1(m_4, m_3, q - m_1, q - m_2)\!+\! R_2 F_1(m_4, m_3, q - m_2, q - m_1)\! \nonumber\\
&+\, R_3 F_2(q - m_3, q - m_1, q - m_2, m_4)+  R_4 F_2(q - m_4, q - m_1, q - m_2, m_3)\qquad\\
&\!\!\!\! +\, (m_1 + m_2 - 2) F_1(q - m_1,  q - m_2, m_3, m_4)\! -\! F_3(q - m_1, q - m_2, m_3, m_4)\bigg]\,.\nonumber
\end{eqnarray}
Here, the first four terms come from the terms in \eqref{fourpoint1}
containing second derivatives of the first-order counterterms, and the last
two terms come from the term with the third derivative of the action
$S^{(0)}$. The overall normalization factor\footnote{Here we suppress usual dimensional factor $v_{*1}^{\delta_{1234}}$.} is
\be
N=\frac{1}{2} c_{1,2}^2 q^{2q-4}\,.
\ee

For the four-point correlators, the universal ratio, which is independent of
the normalizations, can be constructed from \eqref{fourpoint2}, \eqref{Z12},
and \eqref{Z0result}:
\be
\langle\langle O_{m_1n_1}O_{m_2n_2}O_{m_3n_3}O_{m_4n_4} \rangle\rangle_{\text{norm}}=
\frac{Z_{m_1n_1,m_2n_2,m_3n_3,m_4n_4} Z_0}{(\prod_{i=1}^4 Z_{m_in_i,m_in_i})^{\frac{1}{2}}}\,.
\label{4-point MG-dual}
\ee
This result is to be compared with the four-point correlator in MLG
calculated in \cite{BelAlZam} using the standard continuous approach,
\begin{align}
& \frac{ \langle\langle O_{m_1n_1}O_{m_2n_2}O_{m_3n_3}O_{m_4n_4} \rangle\rangle}{
\left( \prod_{i=1}^4 \langle\langle O^2_{m_i,n_i}\rangle\rangle\right)^{\frac{1}{2}} }  =
\frac{\prod_{i=1}^4 |m_ip-n_iq| }{ 2p(p+q)(p-q)}\times \nonumber \\
& \times\left[\sum_{i=2}^4 \sum_{r=-(m_1-1)}^{m_1-1}\sum_{t=-(n_1-1)}^{n_1-1} |(m_i-r)p-(n_i-t)q| -m_1n_1(m_1p+n_1q) \right]\,,
\label{4-point MG}
\end{align}
where $p=q+1$ for the unitary series. 

One further point should be noted.
Expression \eqref{4-point MG} also has some restrictions that should be
taken into account. In particular, the {\it active} field for which the
operator product expansion is used must have the smallest product $m_in_i$
among the four pairs. Moreover, the number of the conformal blocks
\cite{BelAlZam} must be equal to this number. This restriction together with
the requirement that higher counterterms be absent can be satisfied for
general $q$ if we consider symmetric correlation functions of the form
$\langle\langle O_{mn}^4\rangle\rangle_{\text{norm}}$. We note that in this
case, expression \eqref{fourpoint1} (only partially symmetric) becomes
completely symmetric with respect to the permutations of the fields, as it
should.

To give some reference points, we quote a few results for the gravitational
Ising, tricritical Ising, and three-state Potts models corresponding to
$M_{3,4}$, $M_{4,5}$, and $M_{5,6}$. 
Two nontrivial completely symmetric
four-point correlators are the correlator of four spin-density operators
$\sigma=\Phi_{12}$ and the correlator of four energy-density operators
$\epsilon=\Phi_{13}$ dressed by the appropriate Liouville exponential
fields. 
We find that the two expressions \eqref{4-point MG-dual} and
\eqref{4-point MG} give the same results:
\begin{center}
\vspace{10pt}
\begin{tabular}{ l || c | c }
    & $\langle\langle \sigma   \sigma  \sigma  \sigma \rangle\rangle_{\text{norm}}$ & $\langle\langle \epsilon   \epsilon  \epsilon  \epsilon \rangle\rangle_{\text{norm}}$\\[6pt]
   \hline
   \hline\noalign{\smallskip}
   $M_{3,4}$ & $-\frac{1}{7}$&  $\frac{75}{11}$ \\[6pt]
   $M_{4,5}$ & $-\frac{1}{5}$ & $\frac{49}{5}$ \\[6pt]
   $M_{5,6}$ & $-\frac{8}{33}$ & $\frac{162}{11}$
\end{tabular}
\vspace{10pt}
\end{center}
We note that to obtain these results, we must, as we fixed in the very
beginning, take pairs $(m,n)$ with $m\geq n$ in \eqref{4-point MG-dual} and
their reflection images $(q-m,p-n)$ in \eqref{4-point MG}.

\section{Conclusions}
\label{sec:Concl}

We have partially analyzed the three- and four-point correlation functions
using the dual approach to MLG. In the domain where the fusion rules are
satisfied, we found agreement with the results of the continuous approach.

According to the results in \refl{Lemma4} and \refl{Lemma5}, a rigorous
analysis requires much more systematic classification. For three-point
functions \eqref{Z123start}, for example, there exist four possible
regions, depending on which interval contains the parameter $q$, for
example, $m_{12}\leq q\leq m_{23}$, etc., where $m_{ij}=m_i+m_j$ and
$i=1,2,3$. Each region, in turn, contains eight subregions according to the
parities of $m_i-n_i$. For all subregions, we must check whether the
corresponding three-point function is analytic, check the fusion rules,
calculate the second-order counterterms, and finally compare with the
results of the continuous approach.

Even this partial analysis reveals the following problems. As we saw, it
turns out that only a special part of the selection rules can be satisfied
using the resonance transformations. Hence, the selection rules of minimal
models become modified after coupling to Liouville gravity. The nature of
this phenomenon is not yet clear. It is natural to assume that a possible
modification of the method is to require that satisfying {\it this special
part} is a necessary condition. Indeed, the selection rules uniquely fix the
form of the counterterms arising on a given level in the nonphysical region.
On higher levels, these counterterms already enter the expressions for the
correlators in the physical region, i.e., in the region where the fusion
rules are not violated, and must therefore coincide with the results in the
continuous approach. We plan to check this conjecture in the near future.

Another interesting question is to explain the nature of the resonance
transformations from the standpoint of the Frobenius manifold structure. We
believe that answering this question may help in finding a possible
modification of the $A_q$ Frobenius manifold such that this modified version
would be connected to MLG without using the resonance transformations.

\vspace{5mm}

\noindent \textbf{Acknowledgements.}
I am grateful to  A.~Belavin, B.~Dubrovin, and Yu.~Rud for the useful
discussions. I thank Professor K.~Narain for the hospitality during my
visit to ICTP in 2014 and the organizers of the 4th Workshop on Geometric
Correspondences of Gauge Theories at SISSA. The study of the aspects of MLG
theory connected with Frobenius manifolds was supported in part by the
Russian Foundation for Basic Research (Grant No. 13-01-90614). The study of
the form of the resonance transformations and the computation of the
correlation functions was performed with the support of Russian Science
Foundation (Grant No. 14-12-01383).

\vspace{5mm}

\setcounter{section}{0}
\setcounter{equation}{0}
\renewcommand{\theequation}{\thesection.\arabic{equation}}
\setcounter{figure}{0}
\setcounter{table}{0}

\appendix

\renewcommand{\thesection}{A}
\section{ Details of the computations of the structure constants}
\label{sec:StrConst}

We start with a few comments on the multiplication law on the cotangent
bundle in the initial coordinates $u^i$,
\be
d u^i \cdot d u^j =\widetilde{C}^{ij}_k(u) du^k\,.
\ee
On the Frobenius manifolds $A_q$, we can construct so-called
canonical\footnote{We note that there are three natural choices on the
Frobenius manifold, initial, flat, and canonical coordinates, and each
has its own advantages.} coordinates $w^i$ such that the metric is diagonal
(but not constant) in $w^i$. Multiplication of the tangent vectors in the canonical
coordinates has the simple form
\be
\frac{\partial}{\partial w^i}\cdot \frac{\partial}{\partial w^j}=\delta_{ij}  \frac{\partial}{\partial w^i}\,.
\ee
On the cotangent space, if we define
\be
d Q(z)= du_1 z^{q-2}+du_2 z^{q-3}+\dots+du_{q-1}
\ee
using canonical coordinates, then we can easily verify the useful
multiplication law property
\be
d Q(y)\cdot d Q(z)=\frac{Q'(y)dQ(z)-Q'(z)dQ(y)}{y-z}\,.
\label{propA1}
\ee
In the left-hand side of \eqref{propA1}, we have
\be
\sum_{m,n=0}^q  du_{m-1}\cdot du_{n-1} y^{q-m}z^{q-n}\,,
\ee
and in  the right-hand side, we have
\be
\frac{1}{y-z}\sum_{m,n=0}^q(q-m) u_{m-1} du_{n-1} \big[y^{q-m-1}z^{q-n}-z^{q-m-1}y^{q-n}\big]\,.
\ee
The expression in brackets can now be written as
\be
\big[\dots\big]=\begin{cases}
(y-z) y^{q-m-2} z^{q-n} \sum_{k=0}^{n-m-2} \big(\frac{z}{y}\big)^k,\qquad n-m\geq2\,\\
0,\qquad n-m=1\,,\\
(y-z) y^{q-n-1} z^{q-m-1} \sum_{k=0}^{m-n} \big(\frac{z}{y}\big)^k,\qquad n-m\leq0\,.
\end{cases}
\ee
Hence, the right-hand side in \eqref{propA1} becomes
\begin{eqnarray}
&\sum_{m,n=0}^q (q-m) u_{m-1} d u_{n-1}\big[\theta(n-m-2)\sum_{k=0}^{n-m-2} y^{q-m-k-2}z^{q-n+k}-
\nonumber\\
&\theta(m-n)\sum_{k=0}^{m-n} y^{q-n-k-1}z^{q-m+k-1}\big]\,,
\end{eqnarray}
where $\theta(x)=1$ if $x\geq0$ and $0$ otherwise. Collecting the terms
$y^iz^j$ in \eqref{propA1}, we obtain\footnote{I am grateful to Boris Dubrovin 
for the explanation regarding this derivation.} the
answer \cite{VBelavin:2014fs}
\begin{eqnarray}
\widetilde{C}_{i}^{j k}=(q+i-j-k+1) u_{j+k-i-2} \Theta(i,j,k)\,,
\end{eqnarray}
where we introduce the function
\begin{eqnarray}
\Theta(i,j,k)=
\begin{cases} \,\,1 \quad \text{if}\quad j,k\leq i \quad \text{and} \quad j+k > i\,,
\\-1 \quad \text{if}\quad j,k> i \quad \text{and} \quad j+k \leq i+q\,, \\
\,\,0 \quad \text{otherwise}\,, \end{cases}
\label{strconst}
\end{eqnarray}
and $u_{-1}=1,u_0=0$. On the other hand, the metric in flat coordinates is
simple, and lowering an index $\alpha$ is just replacing it with $q-\alpha$.
Hence,
\be
C_{\alpha\beta\gamma} =\frac{\partial v^{q-\alpha}}{\partial u^{i}}
\frac{\partial v^{q-\beta}}{\partial u^{j}}\frac{\partial u^k}{\partial v^{\gamma}} \widetilde{C}_{k}^{ij}\,.
\ee
We can write the expansion in the vicinity $\bf{v}_*$
($\delta\bf{v}=\bf{v}-\bf{v}_*$):
\begin{eqnarray}
\frac{\partial u^k}{\partial v^{\gamma}}({\bf{v}})= {U}_{\gamma}^k+
\delta v^{\rho}{U'}_{\rho\gamma}^k+\dots\,,\\
\frac{\partial v^\gamma}{\partial u^{k}}({\bf{v}})= {V}^{\gamma}_k+
\delta v^{\rho}{V'}_{\rho k}^\gamma +\dots\,,
\end{eqnarray}
where the coefficients can be found in terms of binomial coefficients,
\begin{align}
&{U}_{\gamma}^k=\binom{\frac{\gamma-k+q-2}{2}}{\frac{\gamma+k-q}{2}}\bigg(\frac{v_1}{2}\bigg)^{\frac{\gamma+k-q}{2}} &\text{ if }&\quad\frac{\gamma+k-q}{2}\in \mathbb{N}\,,
\label{U1}
\\
&{V}^{\rho}_j=\frac{2\rho}{q+\rho-j}\binom{q-j-1}{\frac{q-\rho-j}{2}}
\bigg(\!\!-\frac{v_1}{2}\bigg)^{\frac{q-\rho-j}{2}} &\text{ if }&\quad\frac{q-\rho-j}{2}\in \mathbb{N}\,,
\label{U2}
\\
&{U'}_{\alpha\beta}^k=\frac{q-k}{q}\binom{\frac{\alpha+\beta-k-1}{2}}{q-k}\bigg(\frac{v_1}{2}\bigg)^{\frac{k+\alpha+\beta-2q-1}{2}} &\text{if }&\quad
\frac{\!k+\!\alpha+\!\beta-\!2q-\!1}{2}\in \mathbb{N}\,,
\label{U3}
\\
&{V'}^{\alpha}_{\beta k}=-\frac{\alpha}{q}\binom{q-1-k}{\frac{\beta-\alpha-k-1}{2}}
\bigg(\!\!-\frac{v_1}{2}\bigg)^{\frac{\beta-\alpha-k-1}{2}} &\text{if }&\quad
\frac{\beta-\alpha-k-1}{2}\in \mathbb{N}\,.
\label{U4}
\end{align}
In \eqref{U1}--\eqref{U4}, if the conditions are not satisfied, then the
corresponding values are equal to zero.

In particular, from \eqref{U1}, we obtain ${\bf u}({\bf v}_*)$,
\be
u^k=\frac{2q}{k+1}\binom{\frac{2q-k-1}{2}}{\frac{k-1}{2}}
\bigg(\frac{v_1}{2}\bigg)^{\frac{k+1}{2}} \text{ if }\quad\frac{k+1}{2}\in
\mathbb{N}\,.
\ee
In this notation, we have
\begin{align}
&C_{\alpha\beta\gamma}({\bf v}_*)={V}^{q-\alpha}_i{V}^{q-\beta}_j{U}_{\gamma}^k\widetilde{C}_{k}^{ij}\,,\\
&\partial_{\rho} C_{\alpha\beta\gamma}({\bf v}_*)\!=\!\\
&
{V'}^{q-\alpha}_{\rho i}{V}^{q-\beta}_j{U}_{\gamma}^k\widetilde{C}_{k}^{ij}\!+\!{V}^{q-\alpha}_i{V'}^{q-\beta}_{\rho j}{U}_{\gamma}^k\widetilde{C}_{k}^{ij}
\!+\!{V}^{q-\alpha}_i{V}^{q-\beta}_j{U'}_{\rho\gamma}^k\widetilde{C}_{k}^{ij}\!+\!{V}^{q-\alpha}_i{V}^{q-\beta}_j{U}_{\gamma}^k\tilde{C'}_{\rho k}^{ij}\,.\nonumber
\end{align}
Some manipulations with the binomial coefficients give \eqref{strconstflat}
and \eqref{derivstrconst}.

\renewcommand{\thesection}{B}
\section{Some properties of the Jacobi polynomials}
\label{sec:Jacobi}
The polynomials $P_n^{(0,b)}(t)$ satisfy the orthogonality condition
\begin{equation}
\int_{-1}^{1} d t (1+t)^b P_n^{(0,b)}(t)P_m^{(0,b)}(t)=\frac{2^{b+1}}{2n+b+1}\delta_{m,n}.
\label{JacobiScalar}
\end{equation}
In the standard normalization, $P_n^{(0,b)}(1)=1$, and the highest
coefficient is
\begin{equation}
P_n^{(0,b)}(t) =\frac{(b+n+1)_{n}}{n!}\, \bigg(\frac{t}{2}\bigg)^n+\dots\,.
\end{equation}

The Jacobi orthogonal polynomials $P_n^{(0,b)}(t)$ are normalized such that
$P_n^{(0,b)}(1)=1$.

Below, we list properties of the Jacobi polynomials that reveal a more
transparent structure of the second-order counterterms:
\begin{equation}
\begin{aligned}
&\frac{d}{d t} P_n^{(0,b)}(t)=\frac{1}{2}\sum_{k=0}^{n-1} (2k+b+2) P_{k}^{(0,b+1)}(t)\,,  \\
&\frac{d}{d t} \left[(1+t)^{b+1}P_n^{(0,b+1)}(t)\right]=(1+t)^b\sum_{k=0}^{n} (2k+b+1) P_{k}^{(0,b)}(t)\,,
\end{aligned}
\label{JacobiProp1}
\end{equation}
where the second equation can be derived from the first using integration by
parts. Another useful property of the Jacobi polynomials is
\be
\frac{d}{d t} P_n^{(0,b)}(t)=\frac{b+n+1}{2} P_{n-1}^{(1,b+1)}(t)\,.
\label{JacobiProp2}
\ee

\renewcommand{\thesection}{C}
\section{Details of the calculation of the four-point correlator}
\label{sec:App4point}
\subsection*{Third derivatives of the action}
We first discuss the third derivative $S^{(0)}_{\rho\sigma\eta}$. Using the
definition and taking our basic recursion into account, we obtain
\be
\frac{\partial^3 S^{(0)}}{\partial v_{\rho}\partial v_{\alpha}\partial v_{\beta}}=
-\frac{1}{c_{1,2}}\frac{\partial^3 \theta_{1,2}}{\partial v_{\rho}\partial v_{\alpha}\partial v_{\beta}}=
-\frac{1}{c_{1,2}}\bigg(v_1 \frac{\partial}{\partial v_\rho}C_{q-1}^{\alpha\beta}+
C_{q-\rho}^{\alpha\beta}\bigg)\,.
\ee
The WDVV requirement for the structure constants of the Frobenius algebra
(which can be easily verified in our case) gives
\be
\frac{\partial}{\partial v_\alpha}C_{q-1}^{\gamma\beta}=
\frac{\partial}{\partial v_1}C_{q-\gamma}^{\alpha\beta}\,.
\ee
With this result, it is easy to find
\be
\frac{\partial^3 S^{(0)}}{\partial v_{\rho}\partial v_{\alpha}\partial v_{\beta}}=
-\frac{1}{c_{1,2}}\frac{\partial}{\partial v_{1}}
\bigg(v_1 C_{q-\rho}^{\alpha\beta}\bigg)\,.
\ee
Because we are left with only the derivative with respect to $v_1$, we can
now set $v_{k>1}=0$ and use \eqref{strconstflat} for the structure constant.
After lowering the indices, we obtain expression \eqref{thirdderivS0}.

\subsection*{Second derivatives of the first-order counterterms}
We now discuss the second derivative $S^{(i)}_{\alpha\beta}$. We can split
the calculation schematically into two parts. First, we find
\be
\frac{\partial^2 S^{(i)}}{\partial v^\alpha \partial v^\beta}=\sum_{\gamma=1}^{q-1}
C_{\alpha\beta\gamma} \bigg(\!\!-\frac{v_1}{q}\bigg)^{1-\gamma}
\frac{\partial^2 S^{(i)}}{\partial v_1 \partial v^{\gamma}}\,.
\label{2derivS1}
\ee
We then use the same trick as for $S^{(0)}_{\beta\rho\eta}$: we set $v_{k>1}=0$
and use the explicit form of $S^{(i)}_{\gamma}$ in terms of Jacobi polynomials
\eqref{Smn_alpha}. Taking \eqref{JacobiProp2} into account, we obtain
\be
\frac{d}{d t} P_{k}^{(0,b)}(1)=\frac{(b+k+1)k}{2}\,.
\ee
We now prove \eqref{2derivS1}. Using the explicit form of the first-order
counterterms for even $m-n$, we write the expansion in terms of Jacobi
polynomials,
\be\label{Smn0App}
\frac{\partial S^{(mn)}}{\partial v_\alpha} =\delta_{m,\alpha} \widetilde{N}_{mn}
\sum_{k=0}^{\frac{m-n}{2}} b_{\frac{m-n}{2}-k}
\bigg(\!\!-\frac{v_1}{q}\bigg)^{(\frac{m-n}{2}-k)q}\,,
\ee
where $b_k$ are expressed in terms of the coefficients of the Jacobi
polynomials. Here, we prefer to absorb irrelevant factors in \eqref{Smn} in
the normalization $\widetilde{N}_{mn}$. With the results in \refl{Lemma4}
taken into account, it then follows that
\be\label{SmnApp}
S^{(mn)}=\sum_{k=0}^{\frac{m-n}{2}} \widetilde{A}_{m,m-n-2k} \theta_{m,m-n-2k}\,,
\ee
where
\be
\widetilde{A}_{m,m-n-2k} x_{m,m-n-2k}=\widetilde{N}_{mn}  b_{\frac{m-n}{2}-k} \,.
\ee
Differentiating \eqref{SmnApp} twice, we obtain
\begin{align}
\frac{\partial^2 S^{(mn)}}{\partial v_\alpha\partial v_\beta} &=
C_{q-m}^{\alpha\beta} \sum_{k=0}^{\frac{m-n}{2}} \widetilde{A}_{m,m-n-2k}
\frac{\partial\theta_{m,m-n-2k-1}}{\partial v_{q-m}}\nonumber\\
&= C_{q-m}^{\alpha\beta}\bigg(\!\!-\frac{v_1}{q}\bigg)^{m} \widetilde{N}_{mn}
\sum_{k=0}^{\frac{m-n}{2}} b_{\frac{m-n}{2}-k} \frac{y_{m,m-n-2k-1}}{x_{m,m-n-2k}}
\bigg(\!\!-\frac{v_1}{q}\bigg)^{\frac{m-n-2k-2}{2}q}\,.
\end{align}
From \eqref{xy}, we derive
\be\label{xy1App}
\frac{y_{m,m-n-2k-1}}{x_{m,m-n-2k}}=\frac{m-n-2k}{2}\,.
\ee
Hence, on the line $v_{k>1}=0$, we have
\be
\frac{\partial^2 S^{(mn)}}{\partial v_\alpha\partial v_\beta}
= C_{q-m}^{\alpha\beta}\bigg(\!\!-\frac{v_1}{q}\bigg)^{m-q} \widetilde{N}_{mn}
\sum_{k=0}^{\frac{m-n}{2}} b_{\frac{m-n}{2}-k}\bigg(\frac{m-n}{2}-k\bigg)
\bigg(\!\!-\frac{v_1}{q}\bigg)^{\big(\frac{m-n}{2}-k\big)q}\,.
\ee
Comparing this expression with \eqref{Smn0App}, we find
\be
\frac{\partial^2 S^{(mn)}}{\partial v_\alpha\partial v_\beta}
= C_{q-m}^{\alpha\beta} \bigg(\!\!-\frac{v_1}{q}\bigg)^{1-(q-m)}
\frac{\partial^2 S^{(mn)}}{\partial v_1\partial v_m} \,.
\ee
A similar consideration can be performed for odd $m-n$ with the difference
that instead of \eqref{xy1App}, we use
\be
\frac{x_{m,m-n-2k-1}}{y_{m,m-n-2k}}=\frac{m}{q}+\frac{m-n-2k-1}{2}\,.
\ee
The results of both calculations, for both even and odd $m-n$, gives
\eqref{2derivS1}.

\subsection*{Products of two structure constants}
We can rewrite \eqref{F1}, \eqref{F2}, and \eqref{F3} in the forms
\begin{align}
&F_1(\chi,\xi,\mu,\nu)=\!\!\!\!\!\!\!\sum_{\gamma\in R(\chi,\xi)\cap R(\mu,\nu)}\,,
\!\!\!\!\!\!\!\! 1\\
&F_2(\chi,\xi,\mu,\nu)=\!\!\!\!\!\!\!\sum_{\gamma\in R(\chi,\xi)\cap \widetilde{R}(\mu,\nu)}\,,
\!\!\!\!\!\!\!\! 1\\
&F_3(\chi,\xi,\mu,\nu)=\!\!\!\!\!\!\!\sum_{\gamma\in R(\chi,\xi)\cap R(\mu,\nu)}\,,
\!\!\!\!\!\!\!\! \gamma
\end{align}
where
\begin{align}
&R(\chi,\xi)=\big[\chi+\xi-q+1:2:q-|\chi-\xi|-1\big]\,,\\
&\widetilde{R}(\mu,\nu)=\big[\,|\mu-\nu|+1\,\,:2:\,\,2q-\mu-\nu-1\,\big]\,.
\end{align}
Hence,
\begin{align}
&R(\chi,\xi)\cap R(\mu,\nu)=[A_1,B_1]\,,\\
&R(\chi,\xi)\cap \widetilde{R}(\mu,\nu)=[A_2,B_2\,]\,,
\end{align}
where
\begin{align}
&A_1=\text{max}\big(\chi+\xi-q+1,\,\mu+\nu-q+1\big)\,,\\
&B_1=\text{min}\big(q-|\chi-\xi|-1,q-|\mu-\nu|-1\big)\,,\\
&A_2=\text{max}\big(\chi+\xi-q+1,\,|\mu-\nu|+1\big)\,,\\
&B_2=\text{min}\big(q-1-|\chi-\xi|,\,2q-\mu-\nu-1\big)\,,
\end{align}
or
\begin{align}
&A_1=\frac{\chi+\xi+\mu+\nu+2-2q}{2}+\frac{|\chi+\xi-\mu-\nu|}{2}\,,\\
&B_1=\frac{2q-2-|\chi-\xi|-|\mu-\nu|}{2}-\frac{\big||\chi-\xi|-|\mu-\nu|\big|}{2}\,,\\
&A_2=\frac{\chi+\xi-q+2+|\mu-\nu|}{2}+\frac{\big|\chi+\xi-q-|\mu-\nu|\big|}{2}\,,\\
&B_2=\frac{3q-2-|\chi-\xi|-\mu-\nu}{2}-\frac{\big|\mu+\nu-q-|\chi-\xi|\big|}{2}\,.
\end{align}
After replacing $\gamma=A_{1,2}+2s$, we have
\begin{align}
&F_1(\chi,\xi,\mu,\nu)=\sum_{s=0}^{\frac{A_1-B_1}{2}} 1=\frac{B_1-A_1}{2}+1\,,\\
&F_2(\chi,\xi,\mu,\nu)=\sum_{s=0}^{\frac{A_2-B_2}{2}} 1=\frac{B_2-A_2}{2}+1\,,\\
&F_3(\chi,\xi,\mu,\nu)=\sum_{s=0}^{\frac{A_1-B_1}{2}} \big(A_1+2s\big)=\frac{B_1+A_1}{2}\bigg(\frac{B_1-A_1}{2}+1\bigg)\,.
\end{align}
These results lead to \eqref{F1_result}, \eqref{F2_result}, and
\eqref{F3_result}.


\providecommand{\href}[2]{#2}\begingroup\raggedright
\addtolength{\baselineskip}{-3pt} \addtolength{\parskip}{-1pt}

\end{document}